\documentclass[twocolumn, aps, pra, amsmath, amssymb, nofootinbib, superscriptaddress, longbibliography, floatfix, table-of-contents, eqsecnum]{revtex4-2}

\usepackage{nicematrix}
\usepackage{tikz}

\usepackage[]{graphicx}
\usepackage{mathrsfs}
\usepackage[colorlinks, breaklinks]{hyperref}
\usepackage{array}
\usepackage{amsmath}
\usepackage{comment}

\usepackage{type1cm}
\usepackage{lettrine}
\usepackage[english]{babel}
\usepackage{lmodern}
\usepackage{microtype}
\usepackage{booktabs}
\usepackage[T1]{fontenc}
\usepackage[boxed, vlined]{algorithm2e}
\usepackage[margin=0pt, font=small, labelfont=bf, labelsep=endash, justification=centerlast, labelsep=colon]{caption}
\usepackage{braket}
\usepackage[toc,page]{appendix}

\usepackage{amssymb}
\usepackage{mathrsfs}

\usepackage{subfig}
\usepackage{multirow}
\usepackage{etoolbox}
\usepackage{breqn}
\usepackage{float}
\usepackage{bigdelim, rotating}

\makeatletter
\preto\maketitle{%
  \begingroup\lccode`~=`,
  \lowercase{\endgroup
  \let\saved@breqn@active@comma~
  \let~}\active@comma 
}
\appto\maketitle{%
  \begingroup\lccode`~=`,
  \lowercase{\endgroup
  \let~}\saved@breqn@active@comma 
}
\makeatother

\begin{document}

\title{Optical cluster-state generation with unitary averaging}

\author{Deepesh Singh}
\email[]{deepesh.sang@gmail.com}
\affiliation{Centre for Quantum Computation and Communications Technology, School of Mathematics and Physics, The University of Queensland}

\author{Austin P. Lund}
\email[]{a.lund@uq.edu.au}
\affiliation{Dahlem Center for Complex Quantum Systems, Freie Universit\"at Berlin, 14195 Berlin, Germany}
\affiliation{Centre for Quantum Computation and Communications Technology, School of Mathematics and Physics, The University of Queensland}

\author{Peter P. Rohde}
\email[]{dr.rohde@gmail.com}
\homepage[]{www.peterrohde.org}
\affiliation{Centre for Quantum Software \& Information (UTS:QSI), University of Technology Sydney}
\affiliation{Hearne Institute for Theoretical Physics, Department of Physics \& Astronomy, Louisiana State University}

\begin{abstract}
Cluster states are the essential resource used in the implementation of Fusion-based quantum computation (FBQC). We introduce a method to generate high-fidelity optical cluster states by utilising the concept of unitary averaging. This error averaging technique is entirely passive and can be readily incorporated into the proposed PsiQuantum's FBQC architecture. Using postselection and the redundant encoding of Fusion gates, we observe an enhancement in the average fidelity of the output cluster state. We also show an improvement in the linear optical Bell-state measurement (BSM) success probability when the BSM is imperfect.  

\end{abstract}

\maketitle


\section{Introduction}

Quantum computing platforms must inevitably deal with noise.  Achieving perfect isolation from the environment or any implementation imperfections, on a device where one wishes to initialize and read-out data, is likely impossible.  Quantum error correction is a set of methods for which errors can be managed, provided certain guarantees are made on the performance of the components of the computer.  The cost to quantum error correction is the increase in computing resources by way of more qubits and more operations, in order to achieve the same computation.  

Optical quantum computing platforms~\cite{Ralph_2010}, whilst sharing similarities to other platforms through the abstraction of the qubit, have physically different operations and sources of error.  For example, optical systems naturally have access to very large system state sizes through the multitude of modes of propagation available. In fact, a large amount of effort is needed to restrict the systems in which photons evolve into, in order to maximize the quantum interference paths between photons.

The overwhelming drawback of optical quantum computing, is the lack of strong non-linearities (like that of~\cite{Milburn1989_QOFG}).  To achieve quantum computing gates, the output state of a single photon needs to be controlled by the state of another photon.  This description of the interaction with photons, exactly describes a strongly non-linear effect.  One solution to this is the use of off-line resource states coupled with linear evolution and quantum optical detection~\cite{RevModPhys.79.135,PhysRevA.77.062316}.  The non-linear parts of the computation are entirely contained in state preparation and detection, not in the evolution.

There are a number of possible choices within this paradigm.  One particularly promising choice is that of the cluster state based on fusion gates with single photon detection~\cite{PhysRevLett.95.010501}.  A large entangled state (cluster state~\cite{Briegel2001}) is built by `fusing' together smaller entangled states.  When this large entangled state is prepared as a resource, it is consumed by making local measurements in order to progress the computation.

In this paper, we consider the combination of optical based error detection schemes with the Fusion-based quantum computation (FBQC) platform.  Specifically, we combine the redundant encoding of linear scattering matrices over many optical modes considered in~\cite{PhysRevA.97.022324} and~\cite{Vijayan2020robustwstate}.

The redundant encoding scheme is implemented by taking many copies of a desired linearly interacting network and constructing an interferometer that has an interference path that filters out the defects that may be present in any one particular linear network.  This scheme acts to detect continuous errors within device components.  The effect of the construction is to filter the errors and to more likely give a higher quality output than that of using a single interferometer.  The redundant encoding scheme requires more modes, but not more photons.  The encoding scheme also never requires non-linear evolutions. 

The fusion gates of the FBQC platform are a fundamental component and are very commonly utilized.  The standard design for this fusion gate involves a linear interaction followed by photon counting.  To accommodate many fusion gates, many optical modes will be required.  However, these conditions are exactly those for which the redundant encoding scheme operates.  Hence, this method of detecting errors is ideally suited to linear optical devices which may be candidates for the FBQC platform. 


The unitary averaging of the redundant encoding scheme also has some other practical benefits.  It is naturally a passive scheme, removing the need for ancillary photons or circuits, and has no need to perform a feed-forward operation to achieve the error filtering effect. These kinds of benefits will become important in the resource constrained considerations of near-term quantum computing architectures, as was recently proposed for FBQC as published by PsiQ \cite{bartolucci2021fusionbased}. 

In the analysis we present, we use computational symbolic manipulations to present our results.   Given the extremely large number of terms in these expressions, a full presentation of these is not possible.  We therefore present the methodology used to form these manipulations and give results based on the properties of the filtered output state.  This allows us to quantify the improvement that the error filtering effect has.

We have structured this paper as follows: 
In section 2 we give the theoretical background of the redundant encoding process and the FBQC platform.  
Section 3 contains a detailed description of the fusion gate operation that we build upon in the later sections.  
In section 4 we give our application of the encoding to cluster-state generation and present our results on the performance of unitary averaging for fusion gates. 
Section 5 extends the results of section 4 to show the improvement in linear optical Bell-state measurement (BSM) as a result of averaging. 
Finally, we will give some discussion around our resents and present our conclusions.



\section{Background}
In this section we provide the theoretical background of the unitary averaging framework, FBQC, and cluster-state generation using fusion gates. A more detailed description of the fusion gate operation is contained in the next section.

\subsection{Unitary Averaging}
\begin{figure}
    \centering
    \includegraphics[scale=0.1]{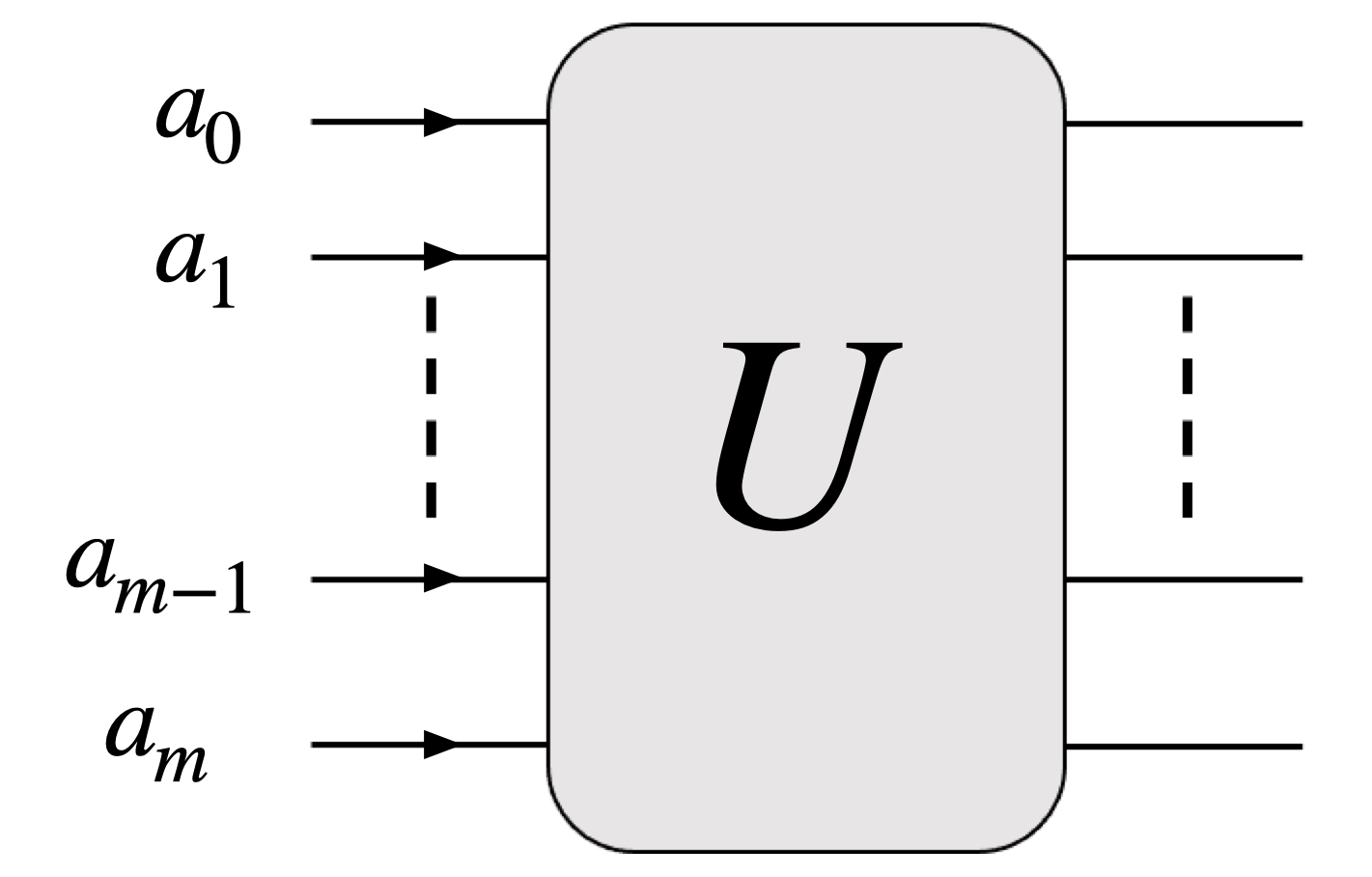}
    \captionof{figure}{Representation of the action of a linear interaction on $m$ input modes bosonic modes represented by their annihilation operators $a_j$.  $U$ represents the matrix $(U_{ij})$ of equation~(\ref{eq:linear}).}
    \label{fig:unitary}
\end{figure}

For any general linear interaction of modes $\mathcal{U}_{U}$, the Heisenberg evolution of the annihilation operators, as shown in Figure~\ref{fig:unitary} is described by: 
\begin{equation}
\label{eq:linear}
  \mathcal{U}_{U}a_{i}\mathcal{U}^{\dagger}_{U} = \sum\limits^{m}_{j=1} U_{ij} a_{j}.
\end{equation}

In experimental settings, it might not always be possible to build the desired unitary $U$ with the required precision. The parameters of $U$ might follow any probability distribution depending on the experimental realisation or fabrication methods. Unitary averaging framework is advantageous in such situations where access to imperfect but multiple unitaries is available. 

\begin{figure}
    \centering
    \includegraphics[scale=0.14]{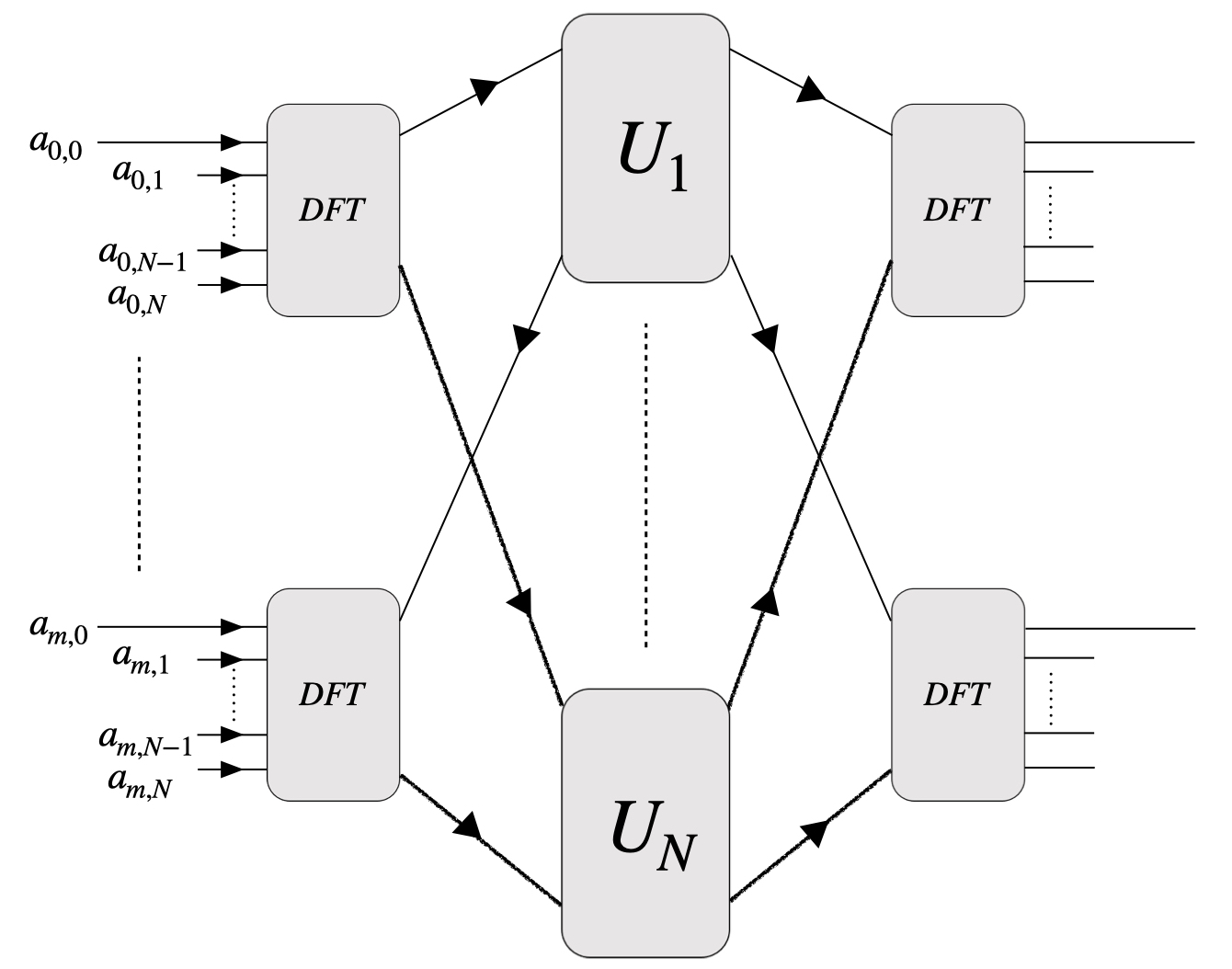}
    \captionof{figure}{Action of an averaged unitary gate on input modes using the representation of Figure~\ref{fig:unitary}.}
    \label{fig:uniavg}
\end{figure}

In the encoding process, each input mode is mixed with $N-1$ ancilla vacuum modes by passing them through a Discrete Fourier Transform (DFT) gate. The output modes of the DFT are related to the input and vacuum modes as:
\begin{equation}
a_{j,r} \longrightarrow \frac{1}{\sqrt{N}} \sum^{N-1}_{k = 0} \omega^{rk} a_{j,k},
\end{equation}
where $\omega=e^{-i 2\pi/N}$ is a primitive $N^{th}$ root of unity. In the notation used, $a_{j,0}$ is the original input mode and $a_{j,i}$, $i \in \{1,2,...,N-1\}$ are the vacuum modes as shown in Figure~\ref{fig:uniavg}. 

The corresponding output modes of the DFTs are then passed through the redundant copies of unitary $U$. The annihilation modes after passing through the $N$ copies, namely $U_{1}, U_{2}, ...,$ and $U_{N}$, evolve as: 
\begin{equation}
a_{j,r} \longrightarrow \sum^{m-1}_{l = 0} (U_{r})_{lj}a_{l,r}
\end{equation}

The modes are then decoded in the end by reapplying the DFT gates, which also follows the evolution described by (2.2). The complete evolution, from encoding, redundant unitary implementation, and decoding can be written as: 
\begin{equation}
a_{j,r} \longrightarrow \frac{1}{N} \sum^{m-1}_{l = 0} \sum^{N-1}_{k',k = 0} (U_{k'})_{lj} \omega^{(r+k)k'}a_{l,k}
\end{equation}

After post-selection on the cases where no photons are present in the output of redundant modes ($k = 0$) the effective evolution of just the original input modes ($r = 0$) is given by: 
\begin{equation}
a_{j,0} \longrightarrow \frac{1}{N} \sum^{m-1}_{l = 0} \sum^{N-1}_{k'=0} (U_{k'})_{lj}a_{l,0},
\end{equation}

which can be rewritten as:
\begin{equation}
a_{j,0} \longrightarrow \sum^{m-1}_{l = 0} (M_{N})_{lj} a_{l,0}
\end{equation}

where, $M_{N} = \frac{1}{N} \sum_{k} U_{k}$. 

Concisely, the relation (2.6) describes the effective evolution of the original modes. In effect, upon successful post-selection of zero photons in the ancilla modes, the action of Figure~\ref{fig:uniavg} reduces to Figure~\ref{fig:unitary} for large $N$ when the parameters of $U_{1}, U_{2}, ...,$ and $U_{N}$ have their mean value equal to the parameters of the desired unitary $U$.  

\subsection{Fusion based quantum computation}
The idea behind FBQC, first introduced in \cite{PhysRevLett.95.010501}, is to perform specific measurements in a certain basis but no particular order on an entangled state constructed by fusing together smaller resource states of a fixed size. 

PsiQ recently proposed an optical implementation of FBQC using dual-rail encoded qubits. The resource states used in the architecture are cluster states such that each qubit $i$ in the state is stabilized by the operator $X_{i}\prod_{j} Z_{j}$, where $j \in$ \{Nearest neighbours of the qubit $i$\}. The final entangled state is created by repeatedly applying Type-II Fusion gates on the outputs of fused resource states.

However, in conjunction with the encoded qubits utilised in PsiQ's architecture, we propose the use of encoded fusion gates as well. Adding ancillary modes with vacuum states would be feasible in PsiQ's integrated photonic circuits and can provide significant improvement in the quality of output cluster states at a low cost.   

\subsection{Cluster state generation}
\begin{figure}
    \centering
    \includegraphics[scale=0.12]{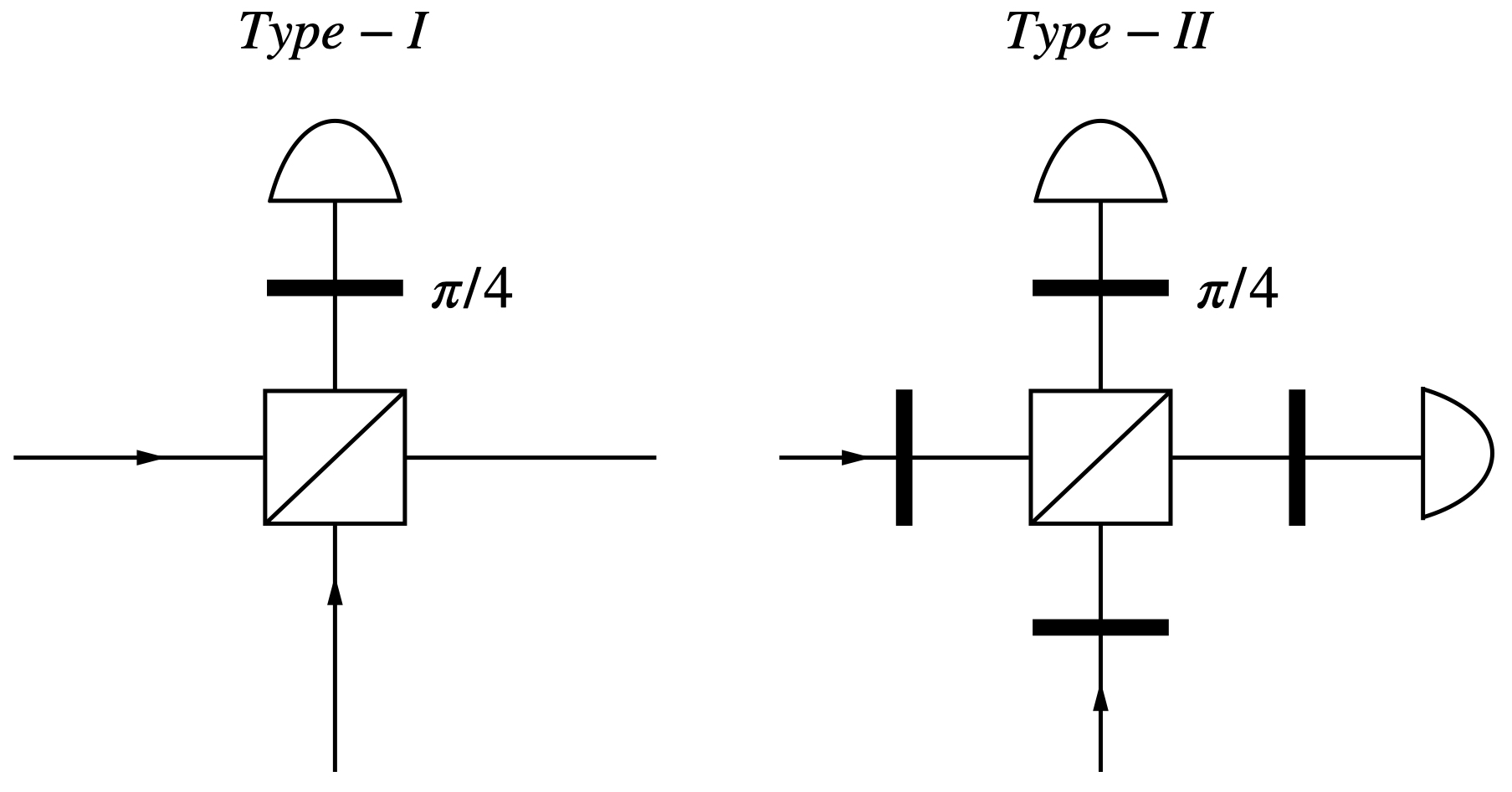}
    \captionof{figure}{The two types of Fusion Gates are demonstrated here. Type-I fusion gates consists of a single polarising beam-splitter (PBS), a $\pi/4$ waveplate on one of the PBS output arms, and a photon-number resolving detector (PNRD). Type-II fusion gates on the other hand, require a single PBS, four $\pi/4$ waveplates, and two polarisation-resolving single-photon (on-off) detectors. }
    \label{fig:fgates}
\end{figure}

We look into the generation of a larger cluster state using Fusion gate operations on two cluster states of smaller length. Although both Type-I and Type-II Fusion gates, shown in Figure~\ref{fig:fgates}, can be used for this purpose, we only highlight the functionality of Type-II Fusion gates for brevity. 

To demonstrate the working of Type-II fusion gates, we use two Bell pairs as our starting resources, which are in turn equivalent to the 2-qubit (or length 2) cluster states $\vert HH \rangle + \vert HV \rangle + \vert VH \rangle - \vert VV \rangle$, and apply the gate on the end qubits of each Bell pair i.e. modes 2 and 3 here. 

The input product state of Bell states $\vert \psi_{in} \rangle$ is given by: 

\begin{align}
\vert \psi_{in} \rangle = \frac{1}{2}(\vert HH \rangle + \vert VV \rangle)_{12}(\vert HH \rangle + \vert VV \rangle)_{34}
\end{align}

After evolution through the first layer of waveplates, PBS, the second layer of wave-plates, and post-selection on single photon measurements on both mode 2 and 3, the output state without any normalisation is: 
\begin{align}
\vert \psi_{out} \rangle = (\vert ++++ \rangle + \vert ---- \rangle)_{1234}
\end{align}

Depending on the parity of the two measured photons, which could either be even (if their polarisations are same) or odd (if their polarisations are different), we can get either of the two following states respectively, each occurring with 25\% probability : 

\begin{align}
\vert \psi^{even}_{out} \rangle = (\vert HH \rangle + \vert VV \rangle)_{14}/\sqrt{2}
\end{align}

\begin{align}
\vert \psi^{odd}_{out} \rangle = (\vert HV \rangle + \vert VH \rangle)_{14}/\sqrt{2}.
\end{align}

Note that for the creation of $\vert \psi^{even}_{out} \rangle$, both $HH$ and $VV$ measurements contribute equally with 12.5\% probability. Similarly, for the creation of $\vert \psi^{odd}_{out} \rangle$, both $HV$ and $VH$ measurements contribute equally with 12.5\% probability.

Upon generalisation, it can be shown that any two linear cluster states of length $n$ and $m$ can be fused together using Type-II gates to create another linear cluster state of length $(n+m-2)$ with 50\% probability. In the case of failures, which happen half the time, the end qubits upon which the fusion was implemented, get destroyed, and we are left with two linear cluster states of lengths $n-1$ and $m-1$ respectively. The process of fusion can then be repeated on these smaller cluster states.   

However, creating larger linear cluster states does not suffice to perform universal quantum computation. The same fusion gates can then also be used to create cluster states with a 2D geometry as shown in \cite{PhysRevLett.95.010501}. 

Nevertheless, in practise, there are a few issues related to the experimental use of type-I fusion gates. Their failure, which happens with 50\% probability, breaks the bond between the end qubit and the remaining cluster, creating issues in scaling. Furthermore, it also requires the use of photon-number resolving detectors (PNRDs) which is not always viable. Type-II fusion gates offer solutions for both these issues and hence we present the averaging over it. The results upon averaging however are the same for Type-1 fusion gates as well.    

\section{General Type-2 Fusion gate}
We prefer to work with dual-rail encoded qubits because of its error-detection property, where both photon loss and photon contamination can be detected by the total photon count in the modes~\cite{https://doi.org/10.48550/arxiv.2106.13825}. Furthermore, we choose the dual spatial mode encoding as done in~\cite{https://doi.org/10.48550/arxiv.2106.13825} which circumvents the need for polarisation resolving detectors.

\begin{figure}[H]
    \centering
    \includegraphics[scale=0.12]{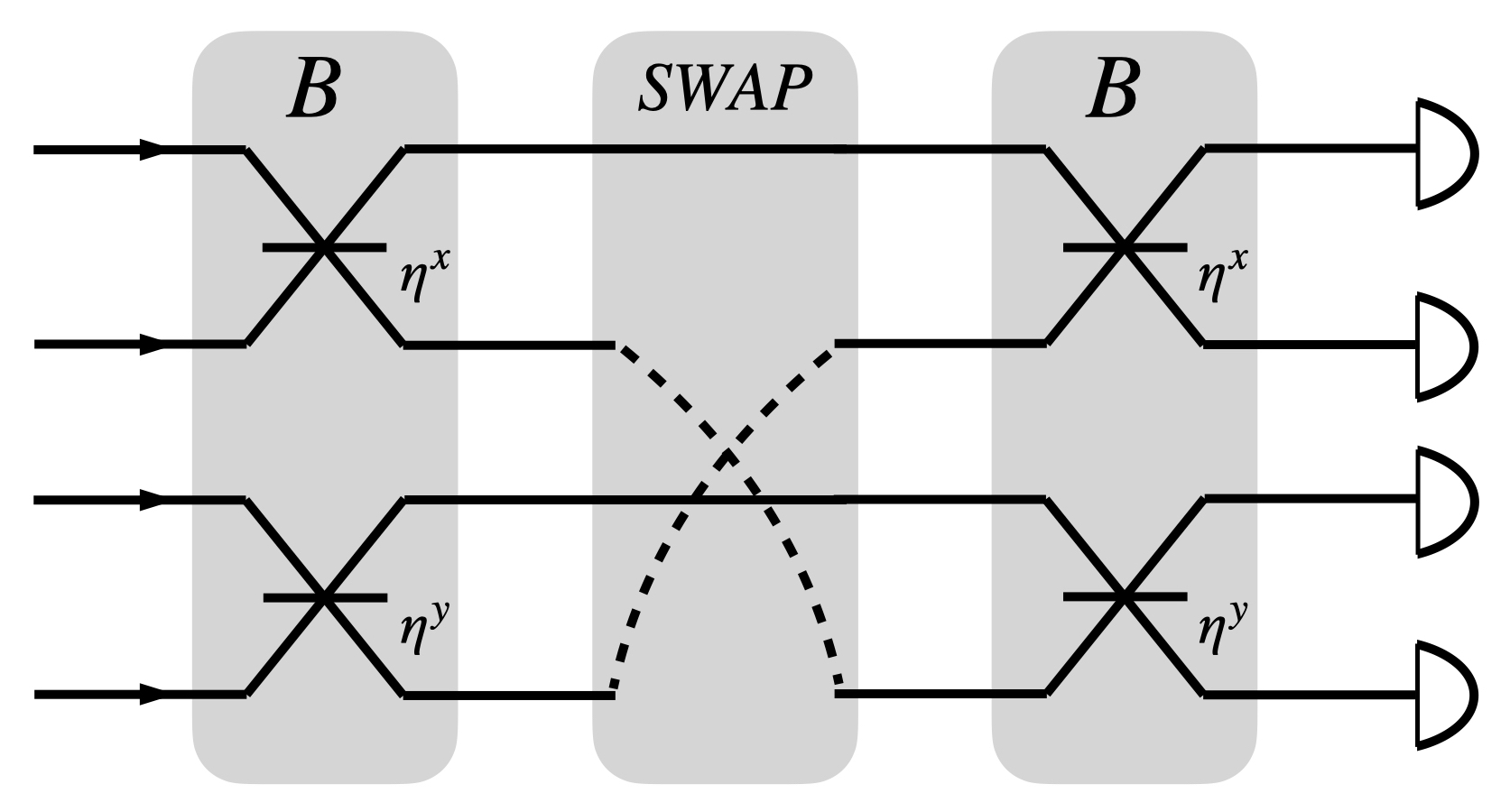}
    \captionof{figure}{Type-II fusion gate in the dual spatial mode encoding. All the beam-splitters have reflectivity 50\% and the dotted lines represent a swapping operation between the second and fourth mode. Note that the basis used to describe this encoding is $H_{1}V_{1}H_{2}V_{2}$ and the gate representation can change for a different basis.}
    \label{fig:type2dual}
\end{figure} 

Within the dual spatial mode encoding of qubits, the wave-plates can be implemented using beam-splitters, and the PBS through swapping operations of certain spatial modes. The Type-II Fusion gate in dual-spatial mode encoding has been shown in Figure~\ref{fig:type2dual}. 

In the following description of Type-II Fusion gate, we assume perfect SWAP gates but erroneous beam-splitters, since SWAP gates are in general easy to implement in the selected encoding. Then, the matrix description of SWAP gate can be written as: 

\begin{center}
   $\text{SWAP}=\left(
    \begin{array}{cccc}
     1 & 0 & 0 & 0 \\
     0 & 0 & 0 & 1 \\
     0 & 0 & 1 & 0 \\
     0 & 1 & 0 & 0 \\
    \end{array}
    \right),$
\end{center}

and for the B matrix, is given by: 

\begin{center}
    $\text{B}=\left(
    \begin{array}{cccc}
     \sqrt{\eta^{x}} & \sqrt{1-\eta^{x}} & 0 & 0 \\
     -\sqrt{1-\eta^{x}} & \sqrt{\eta^{x}} & 0 & 0 \\
     0 & 0 & \sqrt{\eta^{y}} & \sqrt{1-\eta^{y}} \\
     0 & 0 & -\sqrt{1-\eta^{y}} & \sqrt{\eta^{y}} \\
    \end{array}
    \right),$
\end{center}

where the B matrix represents the individual layers of beam-splitters in the fusion gate and reduces to the direct sum of Hadamard matrix, when $\eta^{x}=\eta^{y}=1/2$.

A general Type-2 fusion gate matrix in dual spatial mode encoding in the $(H_{1},V_{1},H_{2},V_{2})$ basis, as represented in Figure~\ref{fig:type2dual}, can then be written as: 
\begin{equation*}
    U(\eta^{x},\eta^{y}) = B(\eta^{x},\eta^{y})*SWAP*B(\eta^{x},\eta^{y}).
\end{equation*}
  
When the beam-splitters are perfect, i.e. $\eta^{x}=\eta^{y}=1/2$, and the application of Type-II Fusion gates on the two Bell pairs creates perfect, i.e. 100\% fidelity, Bell pairs $\vert \psi^{even}_{out} \rangle$ and $\vert \psi^{odd}_{out} \rangle$, each with a 25\% probability i.e. $P_{even}$ and $P_{odd}$ respectively. Therefore, the Type-II fusion gate creates Bell states with a total probability ($P_{single}$) of 50\%, since $P_{single}= P_{even}+P_{odd}$, corresponding to single photon measurements in the output modes corresponding to $(H_{1},V_{1})$ and $(H_{2},V_{2})$.

\begin{figure}
    \centering
    \subfloat[\centering $P_{even}$ as a function of $\eta^{x}$ and $\eta^{y}$]{{\includegraphics[scale=0.58]{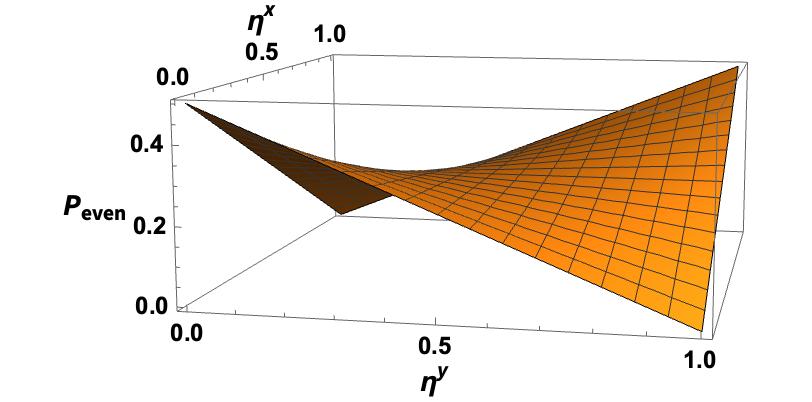} }}%
    \qquad
    \subfloat[\centering $P_{odd}$ as a function of $\eta^{x}$ and $\eta^{y}$]{{\includegraphics[scale=0.58]{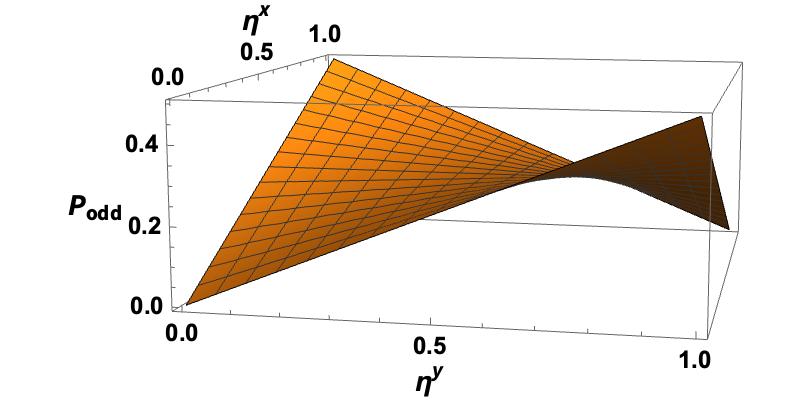} }}%
    \caption{Probabilities of getting even and odd parity photon measurements as a function of the reflectivities of the beam-splitters in the Type-II fusion gate. Note that their sum is constant at 0.5 probability.}%
    \label{fig:example}
\end{figure}

For arbitrary $\eta^{x}$ and $\eta^{y}$, $P_{single}$ remains constant at 50\%. However, the components of $P_{single}$ are a function of $\eta^{x}$ and $\eta^{y}$, as shown in Figure~\ref{fig:example}. Moreover, the description of the fidelity of the corresponding output states is also a function of $\eta^{x}$ and $\eta^{y}$ in general. 

In Table~\ref{tab:mytable}, we consider the different measurement outcomes that contribute to the even and odd parity terms. The relation between their respective probabilities can then be written as:

\begin{align}
    P_{even} &= P_{HH} + P_{VV}, \text{ and} \nonumber \\
    P_{odd} &= P_{HV} + P_{VH}. 
\end{align}

We further see that $P_{HH}=P_{VV}$ and $P_{HV}=P_{VH}$. Therefore Eq. (3.1) reduces to $P_{even}=2P_{HH}=2P_{VV}$ and $P_{odd}=2P_{HV}=2P_{VH}$. 

\begin{table}
  \caption{Two photon measurement outcomes of Type-II Fusion}
  \label{tab:mytable}
  \begin{tabular}{@{}rccccc@{}}
  \toprule
   & Polari- & Success & Output & Target & Normalised \\
   & sations & Prob. & State & Bell State & Fidelity \\ \midrule
\ldelim|{2.2}{*}[\rotatebox{90}{\enspace\bfseries\footnotesize \makebox[0pt]{Even}}] & $HH$ & $P_{HH}$ & $\vert\psi^{HH}\rangle$ & $\vert\psi^{+}\rangle$  & $|\langle \psi^{HH}\vert \psi^{+}\rangle|^{2}/P_{HH}$  \\
 & $VV$ & $P_{VV}$ & $\vert \psi^{VV} \rangle$  & $\vert\psi^{+}\rangle$  & $|\langle \psi^{VV}\vert \psi^{+}\rangle|^{2}/P_{VV}$ \\\addlinespace[1ex]
\ldelim|{2.2}{*}[\rotatebox{90}{\bfseries\footnotesize \makebox[0pt]{\enspace Odd}}] & $HV$ & $P_{HV}$ & $\vert \psi^{HV} \rangle$ & $\vert\phi^{+}\rangle$  & $|\langle \psi^{HV}\vert \phi^{+}\rangle|^{2}/P_{HV}$  \\
  & $VH$ & $P_{VH}$ & $\vert \psi^{VH} \rangle$  & $\vert\phi^{+}\rangle$  & $|\langle \psi^{VH}\vert \phi^{+}\rangle|^{2}/P_{VH}$  \\
   \bottomrule
\end{tabular}
\end{table}

\section{Fusion-gate averaging setup}
\begin{figure}[H]
    \centering
    \subfloat[\centering Explicit description of Type-II Fusion gate averaging over two copies only]{{\includegraphics[scale=0.14]{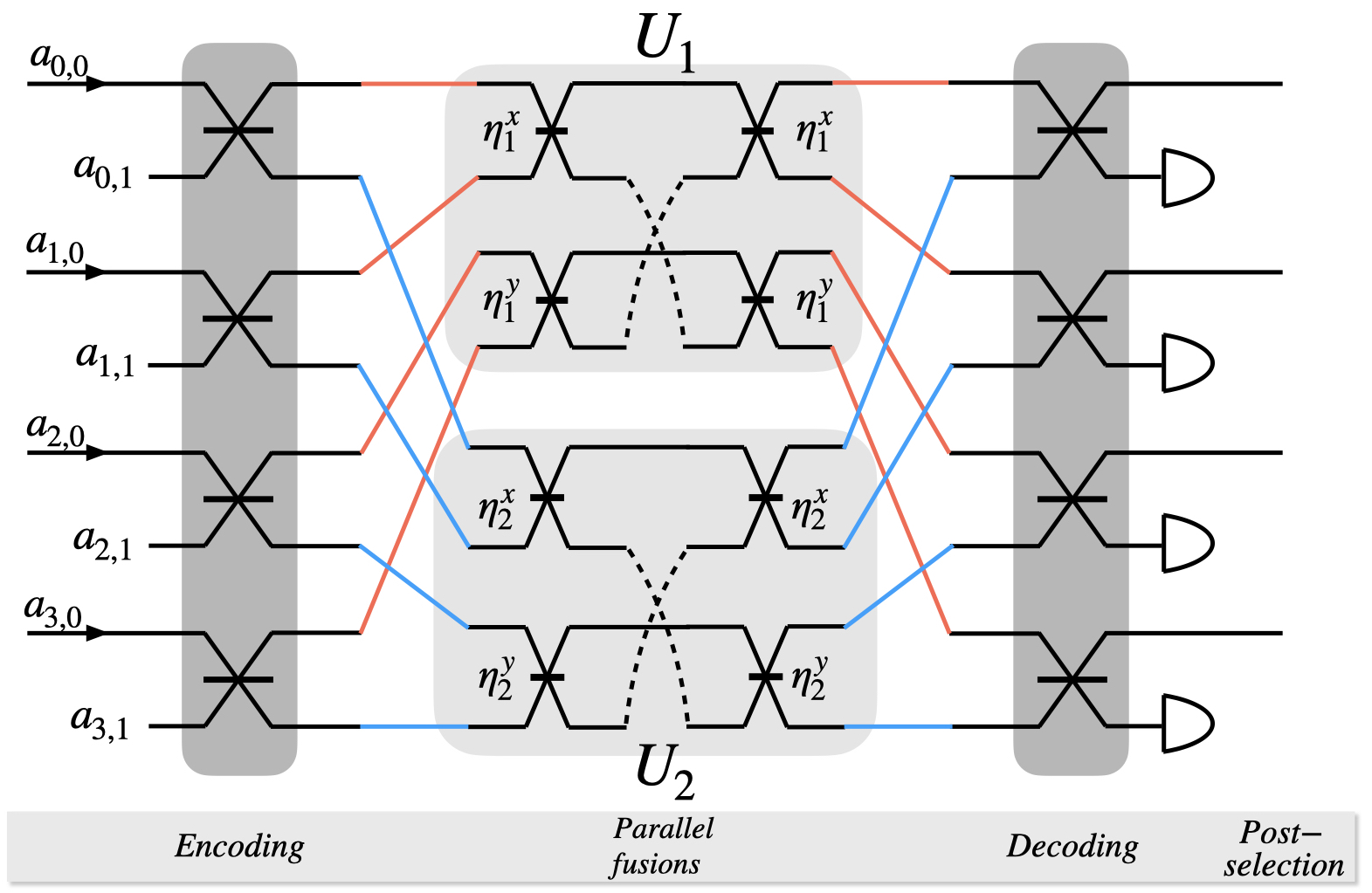} }}%
    \qquad
    \subfloat[\centering General Type-II Fusion gate averaging over N copies]{{\includegraphics[scale=0.14]{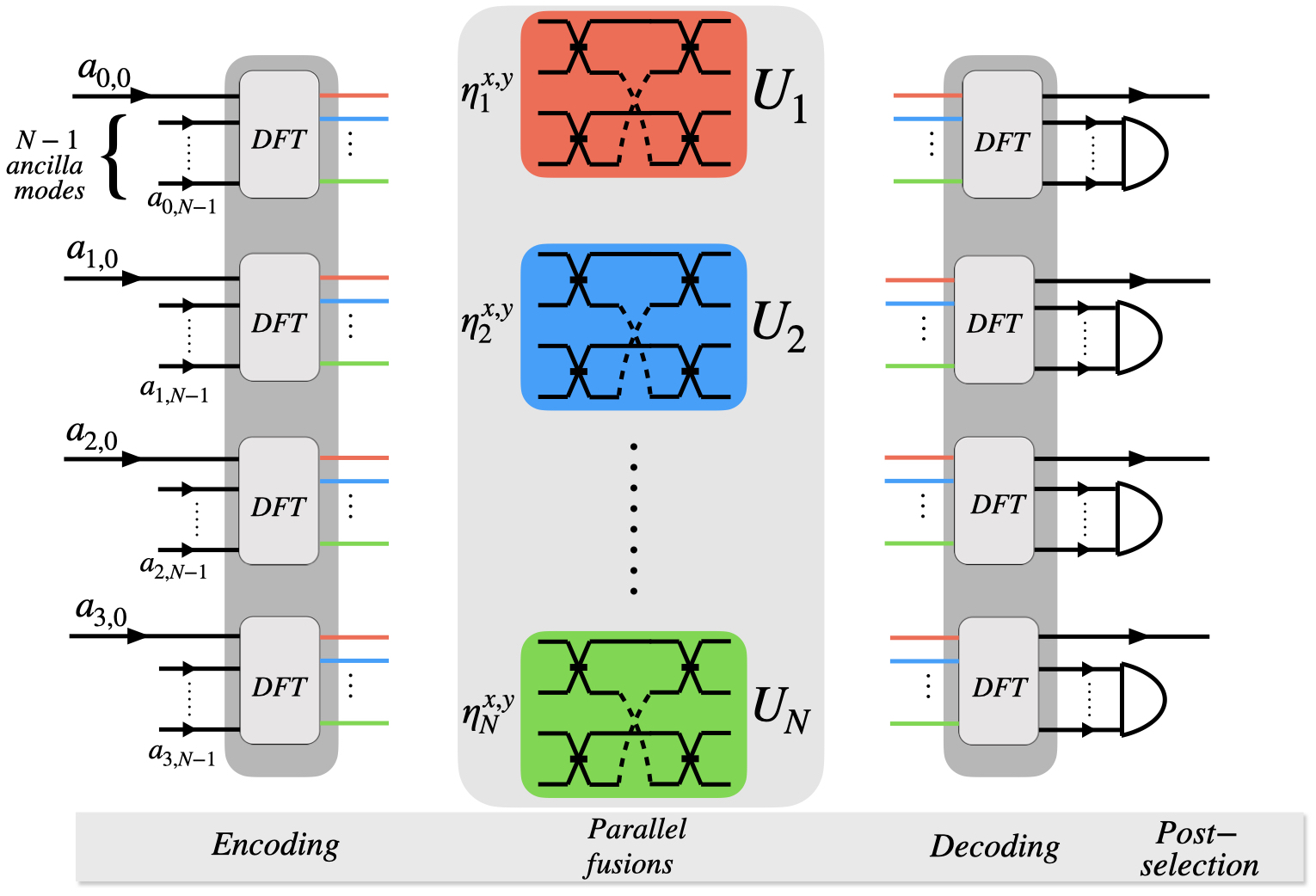} }}%
    \caption{The averaging process essentially consists of four steps: encoding, parallel fusions, decoding, and post-selection on the ancilla modes. After encoding and before the decoding, necessary mode permutations are made to implement the parallel fusion gates, as has been explicitly shown in (a). }%
    \label{fig:fusionavg}
\end{figure}

The elements available for experimental implementations of Type-II Fusion gates might not be perfect. Assuming some probability distribution of their parameters, we can realise the perfect Type-II fusion gates when all the parameters of the contained elements are exactly equal to the mean of their corresponding distribution. If not, the following scheme of averaging, presented in Figure~\ref{fig:fusionavg}, can be implemented to converge to these mean-valued parameters. 

This description of Type-II fusion gate averaging can be extended to arbitrary copies of erroneous fusion gates with the same depth but increasing circuit width as shown in Figure~\ref{fig:fusionavg}(b). In the limit of $N\rightarrow \infty $, as shown in the Section (II.a), the complete circuit tends to the mean of all the parallel fusion gates. 

In this framework, we investigate the effect of the number of redundant fusion gates on the fidelity, success probability, and normalised fidelity of the resulting state. 

For compactness, we only consider the output state when any one of the four possible measurement outcomes is observed. Without loss of generality, we only consider the outputs when both the measurement photons are horizontally polarised i.e. $HH$. As seen from Table~\ref{tab:mytable}, we are interested in the quantities $F^{HH} = |\langle \psi^{HH} \vert \psi^{+} \rangle|^{2}$, $P_{HH}$, and $F^{HH}_{norm}=F^{HH}/P_{HH}$. 

The reflectivities of the beam-splitters belonging to the $i^{th}$ fusion gate $U_{i}$ in Figure~\ref{fig:fusionavg}(b) are represented by $\eta_{i}^{x}$ and $\eta_{i}^{y}$ respectively. For our noise model, we assume that $\eta_{i}^{x}$ and $\eta_{i}^{y}$, $\forall i \in \{1,2,...,N\}$, are independent and identically distributed (iid) random variables. The unitary averaging framework works for all probability distributions owing to the Central Limit Theorem, but the following results assume a Uniform distribution of the variables of the following form: 

\begin{equation}
    \eta_{i}^{z} \sim \text{V}[0.5 - m, 0.5 + m]
\end{equation}

$\forall i \in \{1,2,..,N\} \text{ and } z \in \{ x, y \}$, where $V$ represents the continuous uniform distribution and $m$ is some variable that we sweep over. The Fusion gates are perfect when $m$=0 and show erroneous behaviour as $m \rightarrow \text{0.5} $.

\begin{figure}
    \centering
    \includegraphics[scale=0.35]{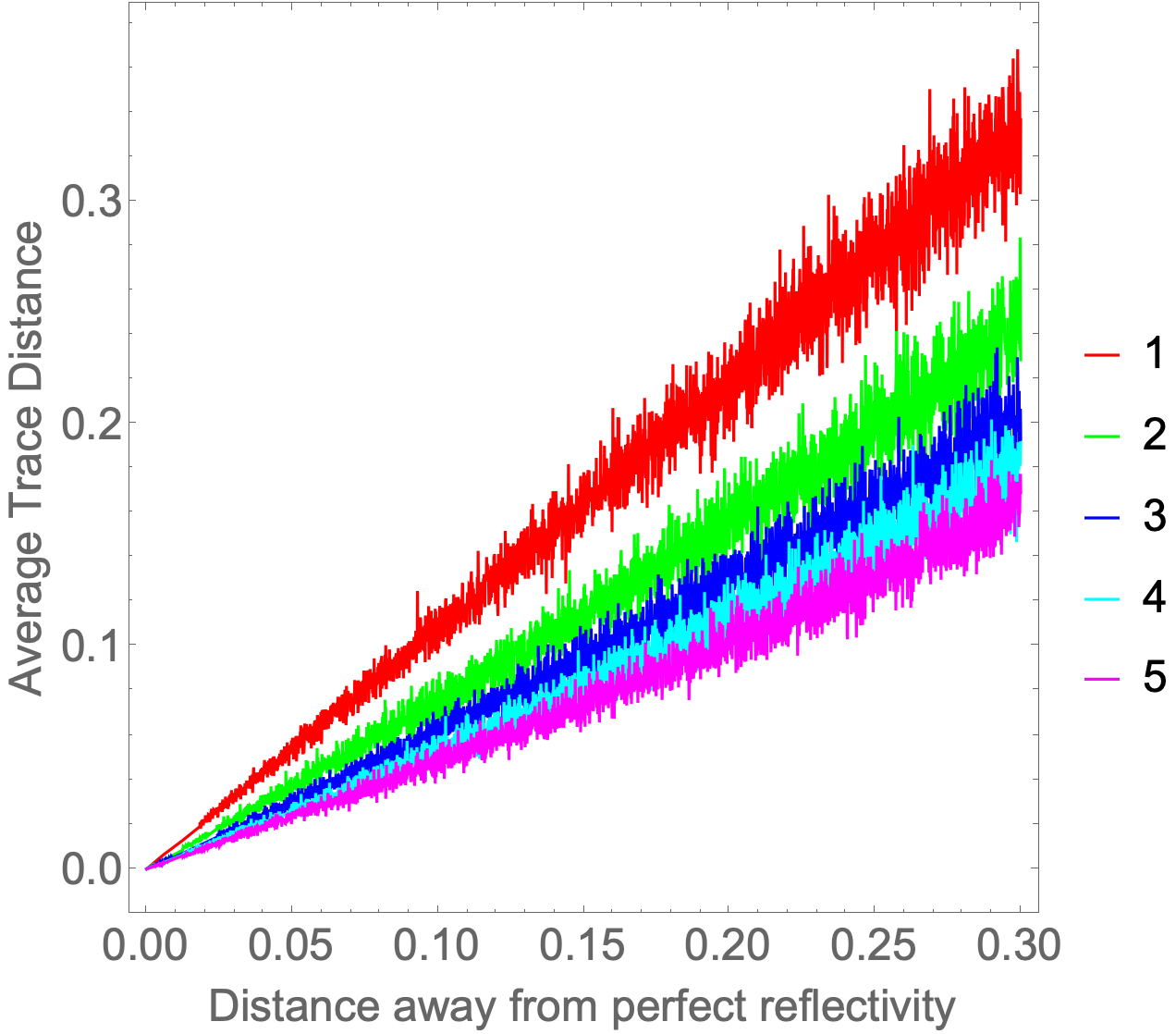}
    \captionof{figure}{Average Trace distance between the perfect fusion gate when both beam-splitters have reflectivity $1/2$ and the average of fusion gates when the BSM reflectivities follow the distribution mentioned in (4.1), as a function of $N$ (number of encoding implemented). The trace distance has been averaged only 50 times which causes the spread in the plot. For averages over larger samples, the average trace distance should converge to different lines for various encoding levels.}
    \label{fig:tracedist}
\end{figure} 

In the present description of the unitary averaging framework, we assume perfect encoding and decoding i.e. the DFT matrices are all perfect. A more exhaustive analysis considering imperfect fusion, encoding, and decoding steps can be done in the future. Investigation of how unitary averaging affects photon loss and photon indistinguishability can also be considered since this will be of importance in experimental implementations of the framework. 

To demonstrate that the average of multiple fusion unitaries with randomly distributed variables is much closer to the perfect fusion gate, we plot the average Trace distance between them as a function of encoding levels in Figure~\ref{fig:tracedist}. For any two matrices $\rho$ and $\sigma$, their trace distance is given by
\begin{align}
\frac{1}{2}\text{Tr}\sqrt{(\rho-\sigma)^{\dagger}(\rho-\sigma)}.
\end{align}
As can be seen in Figure~\ref{fig:tracedist}, on average this distance decreases with the increasing encoding levels, validating the improvement provided by unitary averaging.   

\begin{figure}
    \centering
    \includegraphics[scale=0.35]{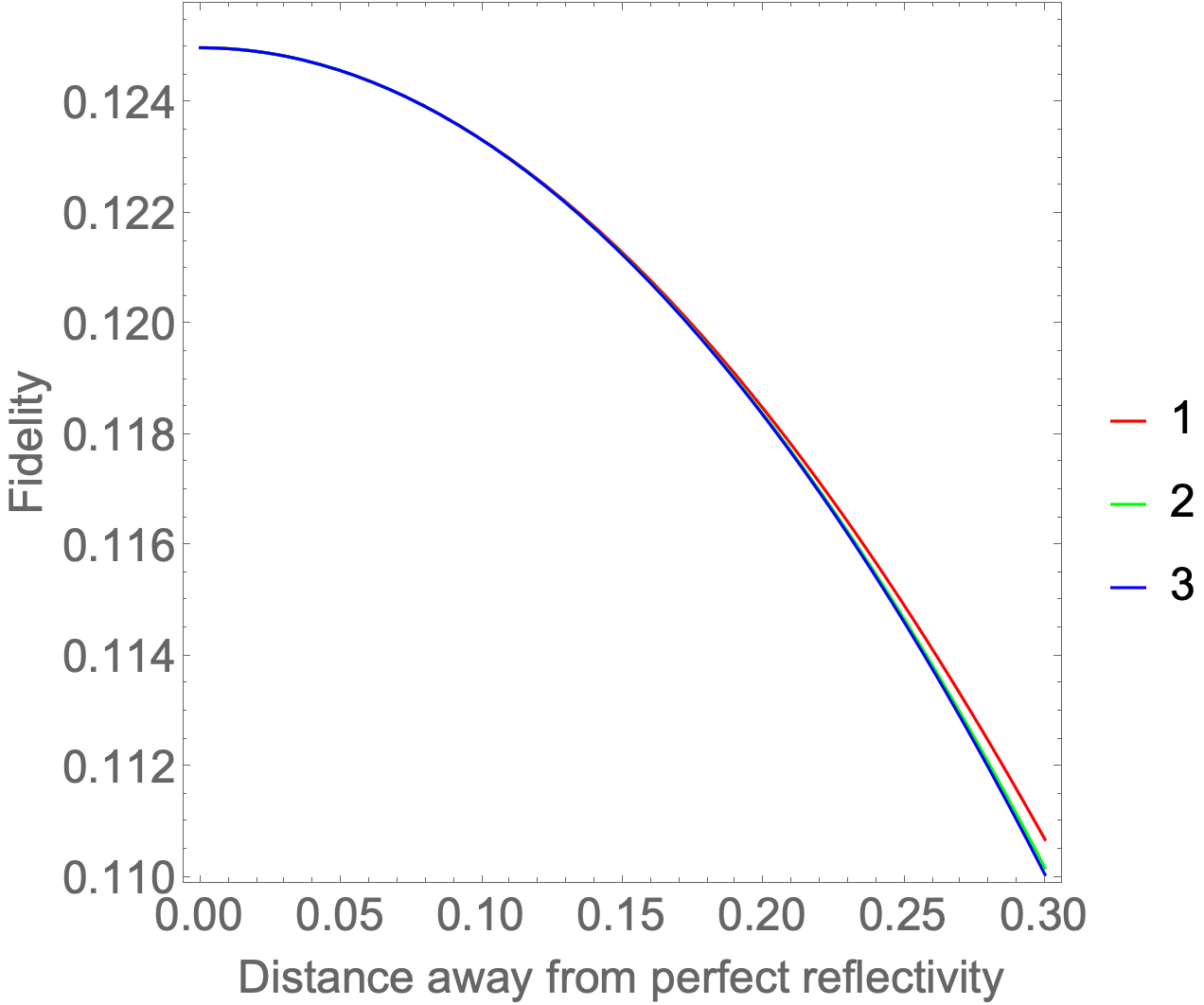}
    \captionof{figure}{Plot of $F^{HH}$, i.e. the overlap between output state $\vert \psi^{HH} \rangle$ and the corresponding target bell state $\vert \phi^{+} \rangle$, as a function of $m$, i.e. the distance away from the mean of the reflectivity uniform distribution as described in (4.1)}
    \label{fig:unnormfid}
\end{figure}

Furthermore, the fidelity without any normalisation, or the overlap between the output state of the fusion gate and the target Bell state, stays almost constant for small imperfections, as a function of the unitary averaging encoding levels as shown in Figure~\ref{fig:unnormfid}. 

\begin{figure}
    \centering
    \subfloat[\centering $P^{HH}$ i.e. the normalisation of the output state $\vert \psi^{HH} \rangle$, as a function of $m$]{{\includegraphics[scale=0.35]{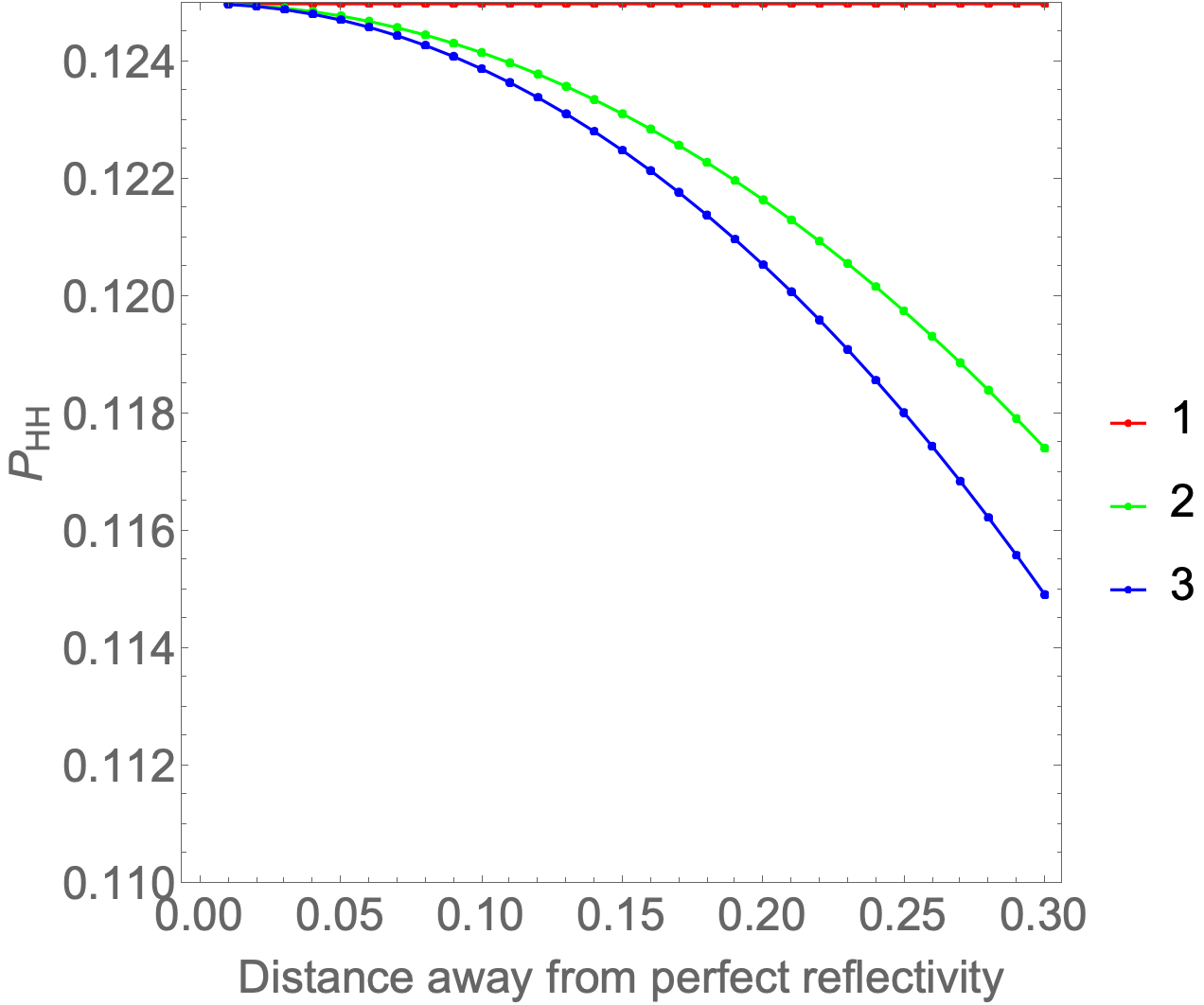}}}%
    \qquad
    \subfloat[\centering $P_{single}$ as a function of $m$]{{\includegraphics[scale=0.35]{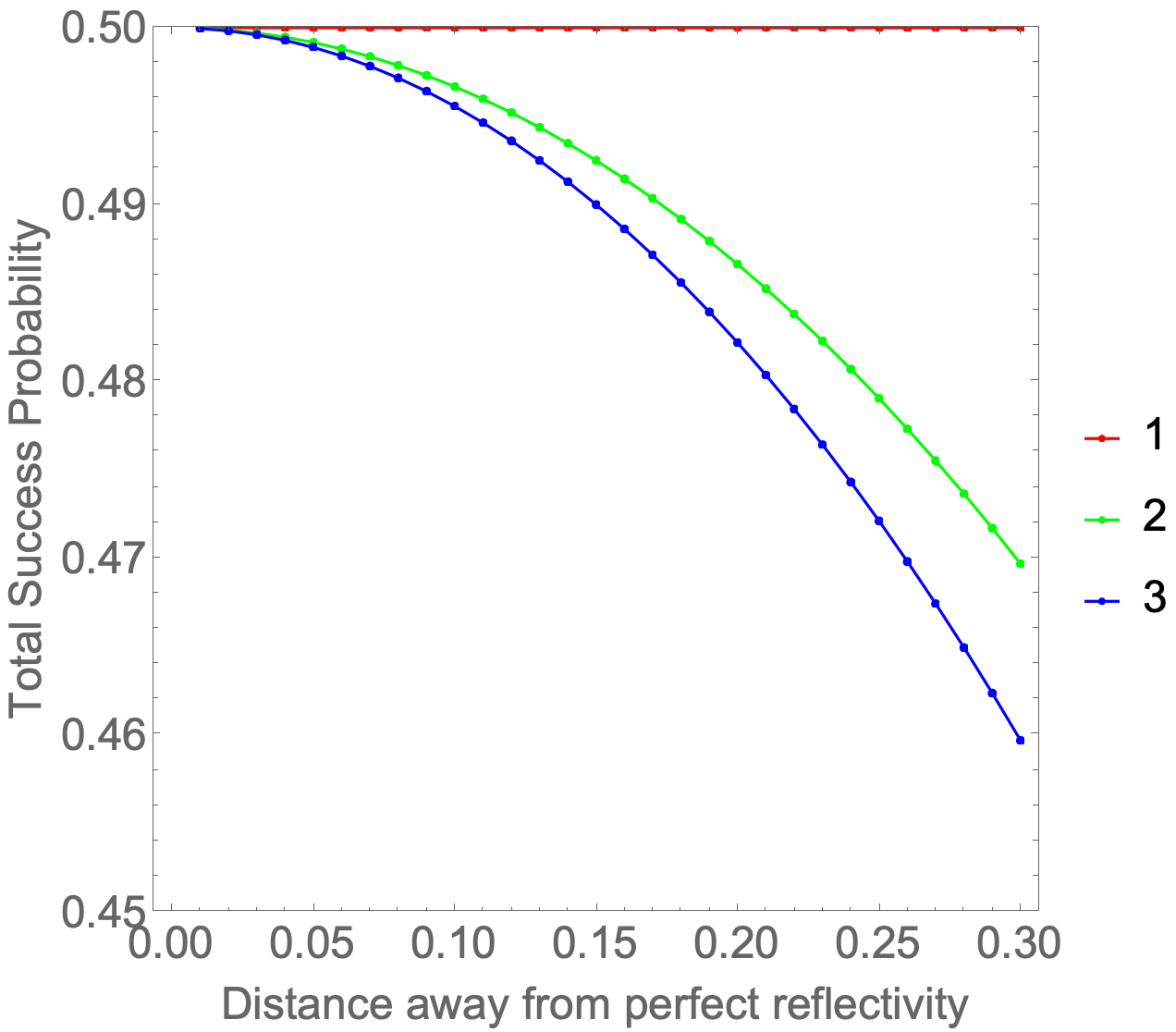}}}%
    \caption{Plots showing change in the success probability of measuring two horizontally polarised photons in (a) and of measuring any two single photons in (b) as a function of $m$}%
    \label{fig:succprob}
\end{figure}

However, the complete success probability of a $HH$ measurement in an averaged fusion gate which includes both successful postselection of no photons in the ancilla modes and successful measurement of two horizontally polarised photons, corresponding to half the even parity measurement probability, decreases with increasing encoding as shown in Figure~\ref{fig:succprob}(a).    

In general, this behavior can be seen for all four possible measurement combinations and not just $HH$. The total success probability (including $HH$, $HV$, $VH$, and $VV$ measurements) of performing an averaged fusion gate as a function of encoding changes as depicted in Figure~\ref{fig:succprob}(b).

\begin{figure}
    \centering
    \includegraphics[scale=0.35]{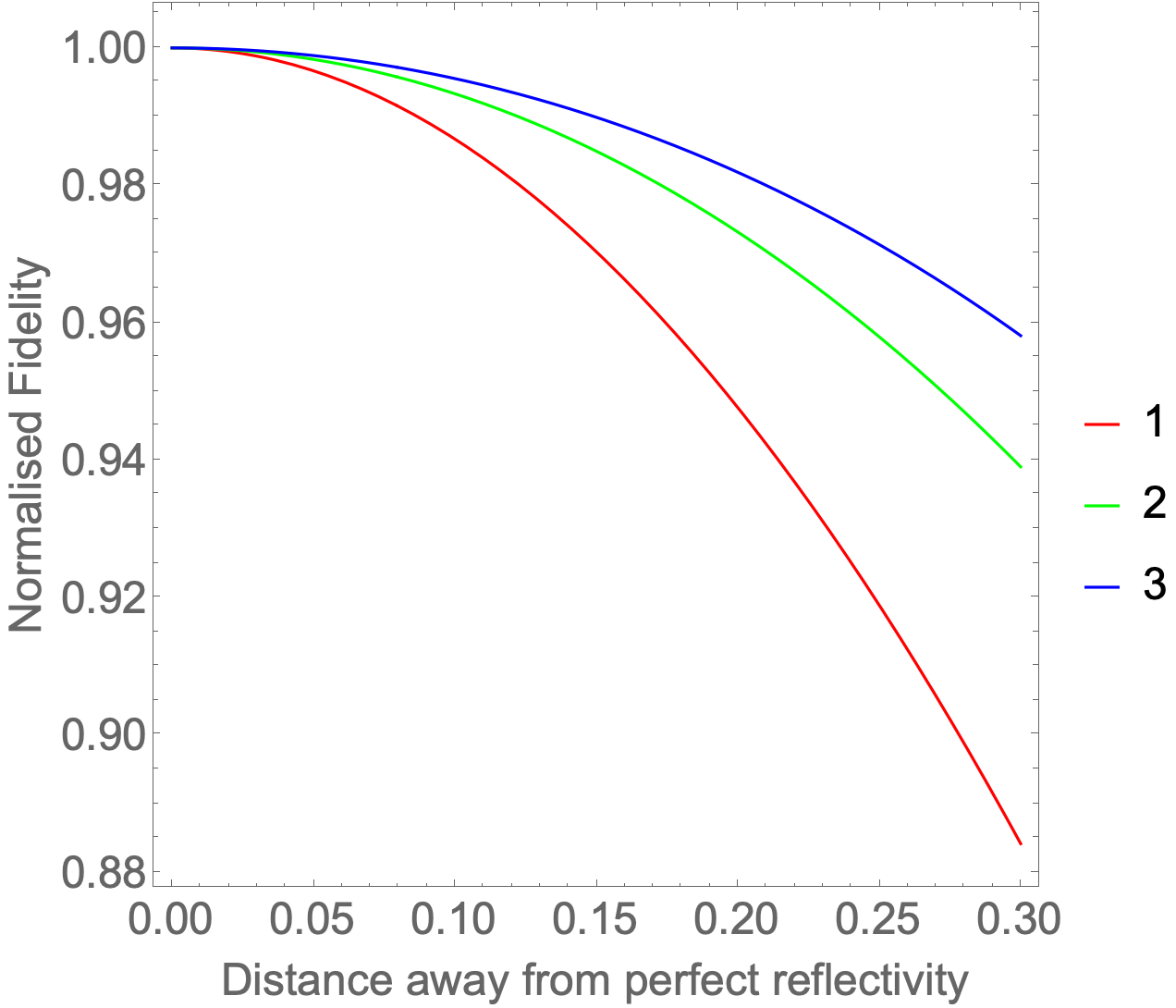}
    \captionof{figure}{Plot of $F^{HH}_{norm}$ as a function of $m$}
    \label{fig:norm_fidelity}
\end{figure}

In effect, the normalised fidelity i.e. the ratio of the fidelity of the output state of an averaged fusion gate and the probability of that output state, increases with the number of encoding $N$ (see Figure~\ref{fig:norm_fidelity}). 

\section{Bell-state measurement averaging setup}

\begin{figure}[H]
    \centering
    \includegraphics[scale=0.2]{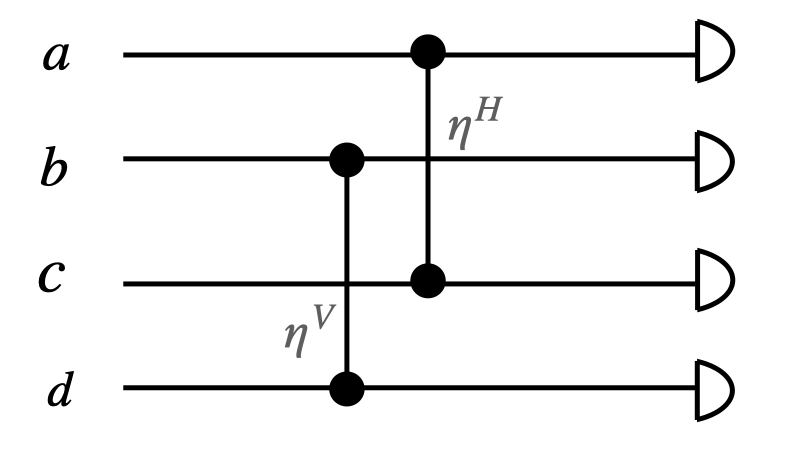}
    \captionof{figure}{A standard Bell-state measurement device in the $H_{1}V_{1}H_{2}V_{2}$ basis. Following the convention used in \cite{https://doi.org/10.48550/arxiv.2106.13825}, the vertical bars represent beam-splitters between the corresponding optical modes and not controlled phase gates.}
    \label{fig:standbsm}
\end{figure}

The standard probabilistic linear optical Bell-state measurement (BSM) device, along with its boosted versions utilising ancillary resources has been described in \cite{PhysRevLett.113.140403,PhysRevA.84.042331,PhysRevLett.114.113603}. In the dual spatial rail encoding for the $H_{1}V_{1}H_{2}V_{2}$ basis, the BSM is shown in Figure~\ref{fig:standbsm}. 

The action of this BSM on the four Bell states is as follows: 

\begin{equation}
    \vert \psi^{+} \rangle = \frac{1}{\sqrt{2}}(\vert 1001 \rangle + \vert 0110 \rangle) \rightarrow \frac{1}{\sqrt{2}} (-\vert 1100 \rangle + \vert 0011 \rangle) 
\end{equation}

\begin{equation}
    \vert \psi^{-} \rangle = \frac{1}{\sqrt{2}}(\vert 1001 \rangle - \vert 0110 \rangle) \rightarrow \frac{1}{\sqrt{2}} (\vert 1001 \rangle - \vert 0110 \rangle)
\end{equation}

\begin{dmath}
    \vert \phi^{+} \rangle = \frac{1}{\sqrt{2}}(\vert 1010 \rangle + \vert 0101 \rangle) \rightarrow \frac{1}{2} (-\vert 2000 \rangle - \vert 0200 \rangle + \vert 0020 \rangle + \vert 0002 \rangle)
\end{dmath}

\begin{dmath}
    \vert \phi^{-} \rangle = \frac{1}{\sqrt{2}}(\vert 1010 \rangle - \vert 0101 \rangle) \rightarrow \frac{1}{2} (-\vert 2000 \rangle + \vert 0200 \rangle + \vert 0020 \rangle - \vert 0002 \rangle)
\end{dmath}

Following the mode naming convention used in Figure~\ref{fig:standbsm}, we can easily represent through Table~\ref{tab:mytable2} the possible measurement outcomes of the four Bell states after the BSM procedure. When the reflectivity of beam-splitters in the BSM is perfect, i.e. $\eta^{H}=\eta^{V}=1/2$, the possible measurement outcomes are shown by the ticks. Note that only the measurement terms corresponding to the $\vert \psi^{+} \rangle$ and $\vert \psi^{-} \rangle$ states are mutually exclusive and hence these two states can always be perfectly discriminated by the BSM.  

\begin{table}
\begin{tabular}{ |c|c|c|c|c|c| } 
\hline
 & $\vert \psi^{+} \rangle$ & $\vert \psi^{-} \rangle$ & $\vert \phi^{+} \rangle$ & $\vert \phi^{-} \rangle$ \\
\hline
$a^{2}$ &   &   & \checkmark  & \checkmark \\ 
$b^{2}$ &  &  & \checkmark  & \checkmark  \\ 
$c^{2}$ &  &  & \checkmark  & \checkmark  \\ 
$d^{2}$ &  &  & \checkmark  & \checkmark  \\
$ab$ & \checkmark  &  &  &  \\ 
$ac$ &  &  & $\times$ & $\times$  \\ 
$ad$ & $\times$ & \checkmark &  &  \\
$bc$ & $\times$ & \checkmark &  &  \\ 
$bd$ &  &  & $\times$ & $\times$ \\
$cd$ & \checkmark  &  &  &  \\
\hline
\end{tabular}
  \caption{Possible measurement combinations for all four Bell states when the BSM is perfect i.e. the reflectivity of both the beam-splitters is 1/2 (represented by ticks only), and imperfect i.e. both the beam-splitters have the same but arbitrary reflectivities (represented by both ticks and crosses).}
    \label{tab:mytable2}
\end{table}

For any $\eta^{H} = \eta^{V} \neq 1/2$, extra measurement combinations are possible and have been represented by the cross symbol. The measurement outcomes of even the $\vert \psi^{+} \rangle$ and $\vert \psi^{-} \rangle$ states do not remain mutually exclusive anymore. Note that the measurement outcomes of the $\vert \psi^{-} \rangle$ Bell state remain unchanged even when both the beam-splitters have the same but arbitrary reflectivity. 

\begin{figure}[H]
    \centering
    \includegraphics[scale=0.35]{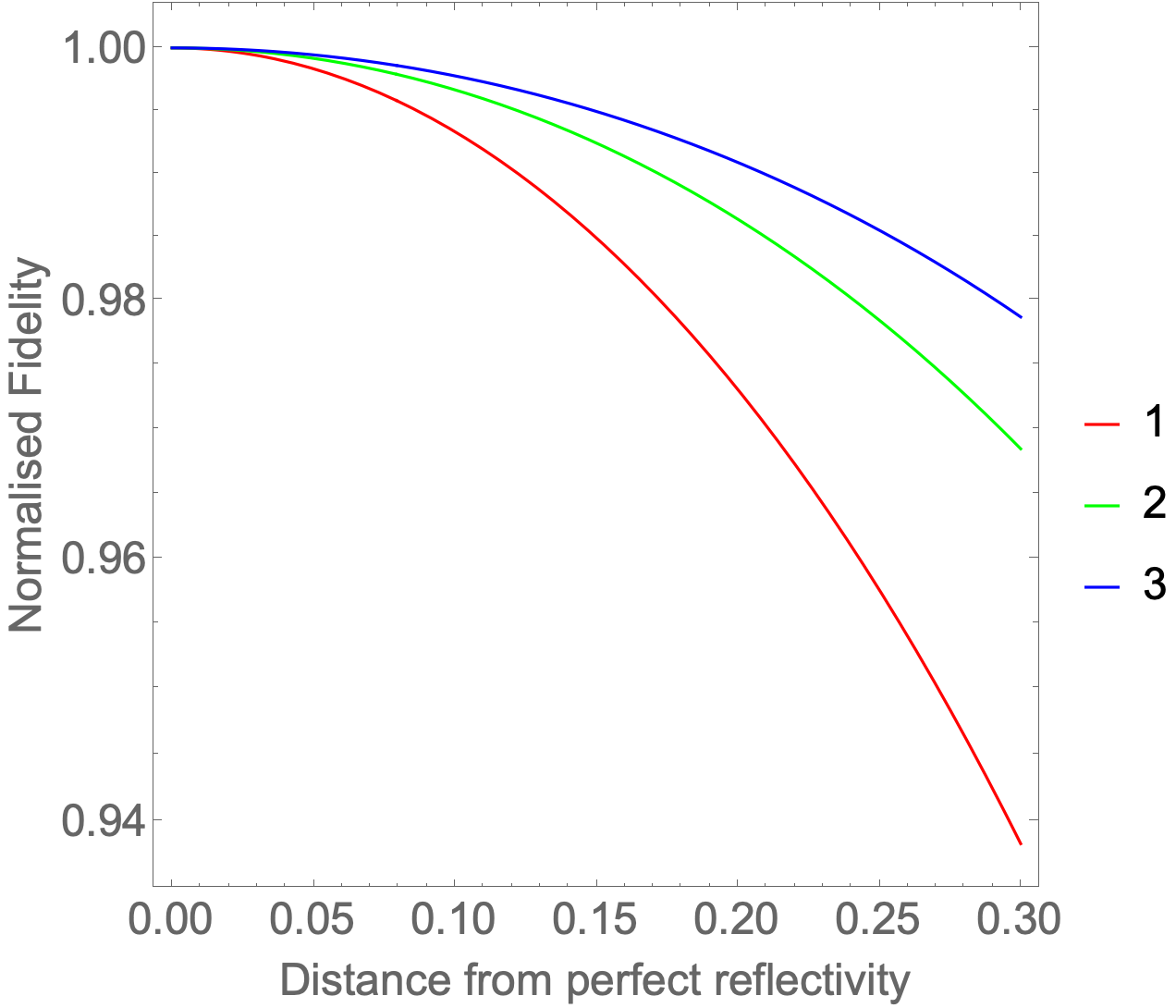}
    \captionof{figure}{Normalised fidelity $F_{norm}$ as a function of $m$}
    \label{fig:bsmfid}
\end{figure}

Since the BSM outputs of $\vert \phi^{+} \rangle$ and $\vert \phi^{-} \rangle$ have the same kets regardless of the beam-splitter reflectivities, we only investigate the effect of BSM averaging on the discrimination of $\vert \psi^{+} \rangle$ and $\vert \psi^{-} \rangle$. Moreover, as $\vert \psi^{-} \rangle$ remains unaffected by the same but arbitrary reflectivities as mentioned above, we can choose to only examine $\vert \psi^{+} \rangle$ under BSM averaging. Using Eq (5.1), we compute the normalised fidelity of the output state as the following: 

\begin{equation}
    F_{norm} = \frac{F}{P_{success}}= \frac{\Bigg| \langle \psi^{+}_{out}(\eta) \vert \frac{\vert 0011 \rangle - \vert 1100 \rangle}{\sqrt{2}} \Bigg|^{2}}{ \Bigg|\langle \psi^{+}_{out}(\eta) \vert \psi^{+}_{out}(\eta) \rangle \Bigg|^{2}}. 
\end{equation}

The general expression for the un-normalised fidelity $F$ and the number of redundant encoding $N$ is straightforward and can be written as: 

\begin{dmath}
    F = \Bigg[ \bigg(\sum\limits_{i=1}^{N} \sqrt{\eta^{H}_{i}} \bigg)\bigg(\sum\limits_{i=1}^{N} \sqrt{1 - \eta_{i}^{V}} \bigg) +  \bigg(\sum\limits_{i=1}^{N} \sqrt{1 - \eta_{i}^{H}} \bigg) \bigg(\sum\limits_{i=1}^{N} \sqrt{\eta_{i}^{V}} \bigg) \Bigg]^{2}. 
\end{dmath}

The normalised Fidelity in Figure~\ref{fig:bsmfid} shows an improvement with increasing number of BSM encoding, i.e. whenever BSM is successful upon post-selection in the averaging framework, the output state is closer to the expected state as compared to the output of a non-averaged BSM. This therefore helps in a better distinction between $\vert \psi^{+} \rangle$ and $\vert \psi^{-} \rangle$.  

\begin{figure}
    \centering
    \includegraphics[scale=0.35]{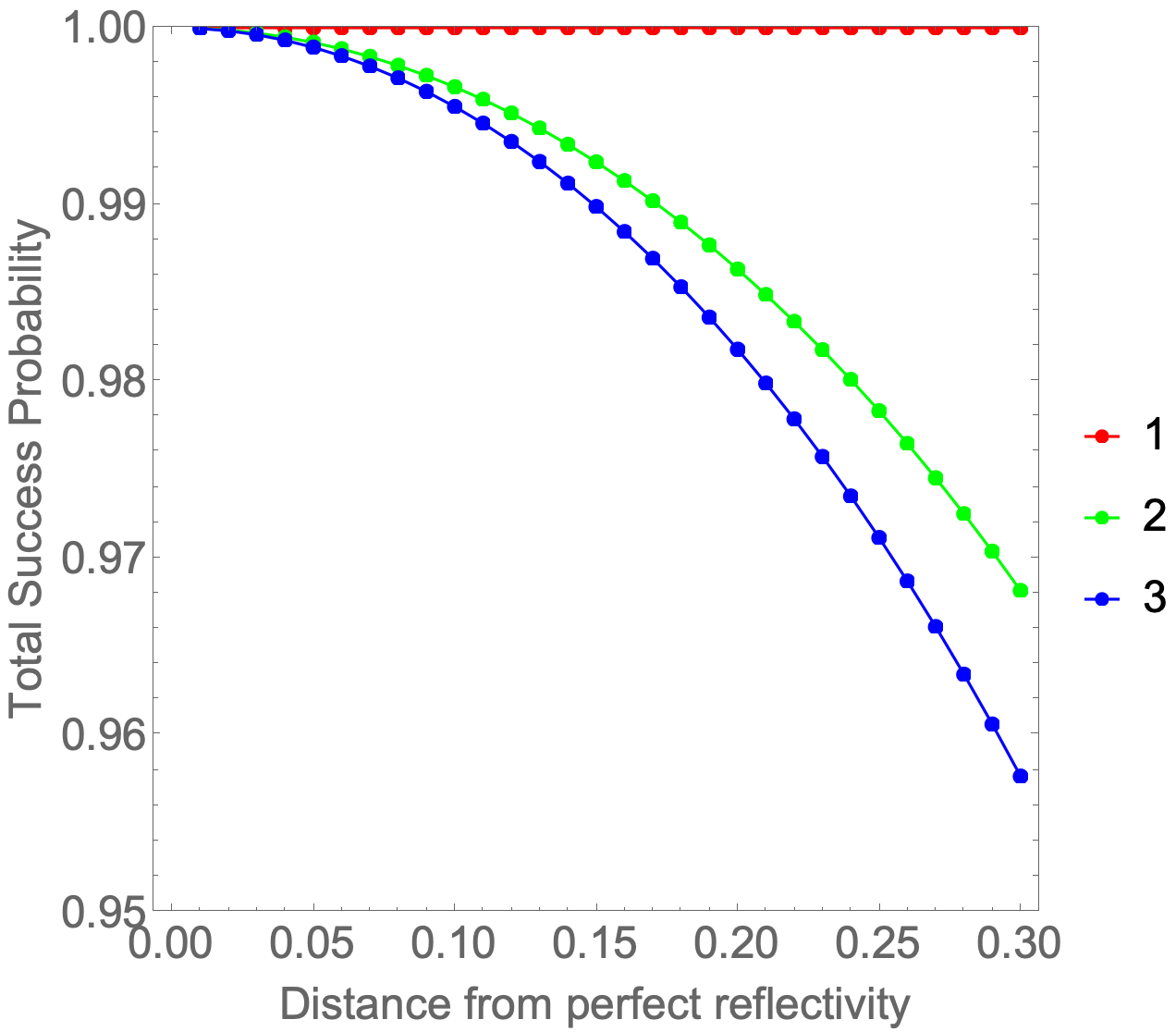}
    \captionof{figure}{$P_{success}$ as a function of $m$}
    \label{fig:bsmsucc}
\end{figure}

For $N$=1, the success probability stays constant at $P_{success}$=1 (see Figure~\ref{fig:bsmsucc}). For $N$>1, $P_{success}\leq 1$ since it also includes the probability of post-selection on zero photons in the ancilla modes. As $\eta\rightarrow\eta_{mean}$, we can observe that a smaller number of photons go to the ancilla modes and hence $P_{success}\rightarrow 1$. 

The general expressions for $P_{success}$ as a function of the number of redundant encoding $N$ becomes complicated. Explicitly, for $N$=2, we can write: 

\begin{dmath}
    P_{success} = \frac{1}{4} \Big(1+\sqrt{1-\eta_{1}^{H}}\sqrt{1-\eta_{2}^{H}}+ \sqrt{\eta_{1}^{H}}\sqrt{\eta_{2}^{H}} \Big) \Big(1+\sqrt{1-\eta_{1}^{V}}\sqrt{1-\eta_{2}^{V}}+ \sqrt{\eta_{1}^{V}}\sqrt{\eta_{2}^{V}} \Big). 
\end{dmath}

$P_{success}$ expressions for up to $N$ = 5 have been included in the appendix.


\section{Conclusion}

We have demonstrated that the combination of redundant error encoding with the construction of fusion gates used in Fusion-Based Quantum Computing (FBQC) can improve the output state fidelity when the operation of the components within devices is not determined to infinite precision. We have given quantitative values for the fidelity improvement if the reflectivities are chosen from a uniform distribution over a fixed range.  Our computations are based on symbolic manipulation of expressions, and this method limits the amount of redundant encoding we can analyse.  We have shown that with three levels of perfect encoding and decoding, a small but always beneficial improvement in the fidelity of the fusion and BSM operations is possible.

\section*{Acknowledgements}

The authors would like to acknowledge fruitful discussions with R.~Marshman. APL acknowledges support from BMBF (QPIC) and the Einstein Research Unit on Quantum Devices. This research was supported by the Australian Research Council (ARC) under the Centre of Excellence for Quantum Computation and Communication Technology (Project No. CE170100012).

\bibliography{references} 

\onecolumngrid
\appendix*
\section{BSM success probabilities for higher encodings}

\begin{itemize}
    \item $\text{For } N=3, $ 
    \begin{dmath}
    P_{success} = \frac{1}{81} \Big[3+2 \Big(\sqrt{1-\eta_{1}^{H}}\sqrt{1-\eta_{2}^{H}} +  \sqrt{1-\eta_{1}^{H}}\sqrt{1-\eta_{3}^{H}}+  \sqrt{1-\eta_{2}^{H}}\sqrt{1-\eta_{3}^{H}} + \sqrt{\eta_{1}^{H}}\sqrt{\eta_{2}^{H}}+\sqrt{\eta_{1}^{H}}\sqrt{\eta_{3}^{H}}+  \sqrt{\eta_{2}^{H}}\sqrt{\eta_{3}^{H}} \Big) \Big]
    \Big[3+2 \Big( \sqrt{1-\eta_{1}^{V}}\sqrt{1-\eta_{2}^{V}} +  \sqrt{1-\eta_{1}^{V}}\sqrt{1-\eta_{3}^{V}}+  \sqrt{1-\eta_{2}^{V}}\sqrt{1-\eta_{3}^{V}} + \sqrt{\eta_{1}^{V}}\sqrt{\eta_{2}^{V}}+\sqrt{\eta_{1}^{V}}\sqrt{\eta_{3}^{V}}+  \sqrt{\eta_{2}^{V}}\sqrt{\eta_{3}^{V}} \Big) \Big]
    \end{dmath}
    \item $\text{For } N=4, $ 
    \begin{dmath} 
    P_{success} = \frac{1}{64} \Big[2+ \sqrt{1-\eta_{1}^{H}}\sqrt{1-\eta_{2}^{H}} + \sqrt{1-\eta_{1}^{H}}\sqrt{1-\eta_{3}^{H}} + \sqrt{1-\eta_{1}^{H}}\sqrt{1-\eta_{4}^{H}} + \sqrt{1-\eta_{2}^{H}}\sqrt{1-\eta_{3}^{H}} + \sqrt{1-\eta_{2}^{H}}\sqrt{1-\eta_{4}^{H}} + \sqrt{1-\eta_{3}^{H}}\sqrt{1-\eta_{4}^{H}} +  \sqrt{\eta_{1}^{H}}\sqrt{\eta_{2}^{H}}+\sqrt{\eta_{1}^{H}}\sqrt{\eta_{3}^{H}}+\sqrt{\eta_{1}^{H}}\sqrt{\eta_{4}^{H}} +  \sqrt{\eta_{2}^{H}}\sqrt{\eta_{3}^{H}} +  \sqrt{\eta_{2}^{H}}\sqrt{\eta_{4}^{H}} +  \sqrt{\eta_{3}^{H}}\sqrt{\eta_{4}^{H}} \Big] 
    \Big[2+ \sqrt{1-\eta_{1}^{V}}\sqrt{1-\eta_{2}^{V}} + \sqrt{1-\eta_{1}^{V}}\sqrt{1-\eta_{3}^{V}} + \sqrt{1-\eta_{1}^{V}}\sqrt{1-\eta_{4}^{V}} + \sqrt{1-\eta_{2}^{V}}\sqrt{1-\eta_{3}^{V}} + \sqrt{1-\eta_{2}^{V}}\sqrt{1-\eta_{4}^{V}} + \sqrt{1-\eta_{3}^{V}}\sqrt{1-\eta_{4}^{V}} +  \sqrt{\eta_{1}^{V}}\sqrt{\eta_{2}^{V}}+\sqrt{\eta_{1}^{V}}\sqrt{\eta_{3}^{V}}+\sqrt{\eta_{1}^{V}}\sqrt{\eta_{4}^{V}} +  \sqrt{\eta_{2}^{V}}\sqrt{\eta_{3}^{V}} +  \sqrt{\eta_{2}^{V}}\sqrt{\eta_{4}^{V}} +  \sqrt{\eta_{3}^{V}}\sqrt{\eta_{4}^{V}} \Big]
    \end{dmath}
    \item $\text{For } N=5, $ 
    \begin{dmath}
    P_{success} = \frac{1}{625} \Big[5+ 2 \Big( \sqrt{1-\eta_{1}^{H}}\sqrt{1-\eta_{2}^{H}} + \sqrt{1-\eta_{1}^{H}}\sqrt{1-\eta_{3}^{H}} + \sqrt{1-\eta_{1}^{H}}\sqrt{1-\eta_{4}^{H}} +
    \sqrt{1-\eta_{1}^{H}}\sqrt{1-\eta_{5}^{H}} +
    \sqrt{1-\eta_{2}^{H}}\sqrt{1-\eta_{3}^{H}} + \sqrt{1-\eta_{2}^{H}}\sqrt{1-\eta_{4}^{H}} +
    \sqrt{1-\eta_{2}^{H}}\sqrt{1-\eta_{5}^{H}} + \sqrt{1-\eta_{3}^{H}}\sqrt{1-\eta_{4}^{H}} + 
    \sqrt{1-\eta_{3}^{H}}\sqrt{1-\eta_{5}^{H}} +
    \sqrt{1-\eta_{4}^{H}}\sqrt{1-\eta_{5}^{H}} + \sqrt{\eta_{1}^{H}}\sqrt{\eta_{2}^{H}}+
    \sqrt{\eta_{1}^{H}}\sqrt{\eta_{3}^{H}}+\sqrt{\eta_{1}^{H}}\sqrt{\eta_{4}^{H}} +\sqrt{\eta_{1}^{H}}\sqrt{\eta_{5}^{H}} +  \sqrt{\eta_{2}^{H}}\sqrt{\eta_{3}^{H}} +  \sqrt{\eta_{2}^{H}}\sqrt{\eta_{4}^{H}} +
    \sqrt{\eta_{2}^{H}}\sqrt{\eta_{5}^{H}} + \sqrt{\eta_{3}^{H}}\sqrt{\eta_{4}^{H}} + \sqrt{\eta_{4}^{H}}\sqrt{\eta_{5}^{H}} + \sqrt{\eta_{4}^{H}}\sqrt{\eta_{5}^{H}}\Big) \Big] 
    \Big[5+ 2\Big( \sqrt{1-\eta_{1}^{V}}\sqrt{1-\eta_{2}^{V}} + \sqrt{1-\eta_{1}^{V}}\sqrt{1-\eta_{3}^{V}} + \sqrt{1-\eta_{1}^{V}}\sqrt{1-\eta_{4}^{V}} +
    \sqrt{1-\eta_{1}^{V}}\sqrt{1-\eta_{5}^{V}} + \sqrt{1-\eta_{2}^{V}}\sqrt{1-\eta_{3}^{V}} + \sqrt{1-\eta_{2}^{V}}\sqrt{1-\eta_{4}^{V}} +
    \sqrt{1-\eta_{2}^{V}}\sqrt{1-\eta_{5}^{V}} +
    \sqrt{1-\eta_{3}^{V}}\sqrt{1-\eta_{4}^{V}} + 
    \sqrt{1-\eta_{3}^{V}}\sqrt{1-\eta_{5}^{V}} +
    \sqrt{1-\eta_{4}^{V}}\sqrt{1-\eta_{5}^{V}} +
    \sqrt{\eta_{1}^{V}}\sqrt{\eta_{2}^{V}}+\sqrt{\eta_{1}^{V}}\sqrt{\eta_{3}^{V}}+\sqrt{\eta_{1}^{V}}\sqrt{\eta_{4}^{V}}+\sqrt{\eta_{1}^{V}}\sqrt{\eta_{5}^{V}} +  \sqrt{\eta_{2}^{V}}\sqrt{\eta_{3}^{V}} +  \sqrt{\eta_{2}^{V}}\sqrt{\eta_{4}^{V}}+  \sqrt{\eta_{2}^{V}}\sqrt{\eta_{5}^{V}} +  \sqrt{\eta_{3}^{V}}\sqrt{\eta_{4}^{V}} +  \sqrt{\eta_{3}^{V}}\sqrt{\eta_{5}^{V}} +  \sqrt{\eta_{4}^{V}}\sqrt{\eta_{5}^{V}} \Big) \Big]
    \end{dmath}
\end{itemize}

\end{document}